\documentclass[numberedappendix]{emulateapj}
\usepackage{natbib}
\slugcomment{Submitted to ApJ}

\shorttitle{Dry Tidal Features}
\shortauthors{Kawata et al.}

\begin{document}


\title{Are Red Tidal Features Unequivocal Signatures of Major Dry Mergers?}

\author{Daisuke. Kawata\altaffilmark{1,2},
John S. Mulchaey\altaffilmark{1}, 
Brad K. Gibson\altaffilmark{3}, and
Patricia S\'anchez-Bl\'azquez\altaffilmark{4}
}

\altaffiltext{1}{The Observatories of the Carnegie Institution of Washington,
 813 Santa Barbara Street, Pasadena, CA 91101
\email{dkawata@ociw.edu}}
\altaffiltext{2}{
 Swinburne University of Technology, Hawthorn VIC 3122, Australia
}
\altaffiltext{3}{
Centre for Astrophysics, University of Central Lancashire, Preston,
PR1 2HE, United Kingdom
}
\altaffiltext{4}{
Laboratoire d'Astrophysique, \'Ecole Polytechnique F\'ed\'erale de
Lausanne (EPFL) Observatoire, CH-1290, Sauverny, Switzerland
}

\begin{abstract}
We use a cosmological numerical simulation to study the tidal features produced 
by a minor merger with 
an elliptical galaxy.
We find that the simulated tidal features are quantitatively similar to
the red tidal features, i.e., dry tidal features, 
recently found in deep images of
elliptical galaxies at intermediate redshifts.
The minor merger in our simulation
does not trigger star formation due to
active galactic nuclei heating. Therefore, both the tidal features
and the host galaxy are red, i.e. a dry minor merger.
The stellar mass of the infalling satellite galaxy is about $10^{10}$
M$_{\sun}$, and the tidal debris reach the surface brightness
of $\mu_R\sim$27 mag arcsec$^{-2}$. Thus, we conclude that
tidal debris from minor mergers can explain the observed dry
tidal features in ellipticals at intermediate redshifts, although 
other mechanisms (such as major dry mergers) may also 
be important. 
\end{abstract}


\section{Introduction}
\label{sec-intro}

Recently \citet[][hereafter vD05]{vd05} 
analyzed unprecedentedly deep image
of red galaxies at redshift around z$\sim0.1$, 
and found that they often have diffuse
tidal features over the scale of $\sim50$ kpc. Furthermore, there is no
associated star formation with these tidal features. vD05 call such features
``dry tidal features''. vD05 also find that 30 \% of the
red galaxies in their sample are undergoing {\it ongoing} mergers,
and the luminosity difference of these merging galaxies
are often small (the ratio of the luminosities 
between the bigger and smaller galaxies is more than 0.3). 
However, it is still an open question whether most dry tidal
features are the remnants
of such major dry mergers or whether other mechanisms can produce
these features\footnote{Throughout this paper, we call a merger
whose luminosity ratio is greater than 0.3 ``major merger'', and
a merger whose luminosity ratio is less than 0.3 ``minor merger''.}.

Here, we show that deep images of simulated 
red galaxies also display red tidal features. 
The galaxy studied is taken from a fully self-consistent $\Lambda$-dominated
cold dark matter ($\Lambda$CDM) cosmological simulation.
We identify the responsible incident for the dry tidal feature,
and find that the feature is due to 
the tidal tails created from an infall of a relatively 
small satellite galaxy, i.e.\ 
a minor merger. Our simulation suggests that
such minor mergers may be responsible for some of
the dry tidal features observed in vD05.

Our numerical simulations and the 
simulated elliptical galaxy are described in the next section.
Section \ref{sec-dftg} presents a quantitative comparison
between the dry tidal features in the simulated galaxy
and those observed in vD05, and shows that 
the simulated features are consistent with the observed ones.
The implications of this finding and our conclusions are given in the final section.

\section{Target Simulated Elliptical Galaxy}
\label{sec-meth}
The simulated images considered here are based on
the elliptical galaxy model 2
presented in \citet[][hereafter KG05]{kg05}. 
KG05 carried out $\Lambda$CDM cosmological simulations
using the galactic chemodynamics code, {\tt GCD+} \citep{kg03a}.
{\tt GCD+} is a three-dimensional tree $N$-body/smoothed
particle hydrodynamics (SPH) code which incorporates self-gravity,
hydrodynamics, radiative cooling, star formation, supernovae (SNe)
feedback, and metal enrichment. {\tt GCD+} takes account of the chemical
enrichment by both Type~II (SNe~II) and Type~Ia (SNe~Ia) SNe, mass-loss
from intermediate mass stars, and follows the chemical enrichment history
of both the stellar and gas components of the system. 

The cosmological simulation adopts
a $\Lambda$CDM cosmology ($\Omega_0$=0.3, $\Lambda_0$=0.7,
$\Omega_{\rm b}$=0.019$h^{-2}$, $h$=0.7, and $\sigma_8$=0.9),
and uses a multi-resolution technique to achieve high-resolution
in the regions of interest, including the tidal forces from
neighboring large-scale structures.
The initial conditions for the simulations are constructed
using the public software {\tt GRAFIC2} \citep{eb01}. Gas dynamics and
star formation are included only within the relevant high-resolution
region ($\sim$12~Mpc at $z$=0); the surrounding low-resolution region
($\sim$43~Mpc) contributes to the high-resolution region only through
gravity.
 Consequently, the initial condition consists of a total of 190093 dark matter
particles and 134336 gas particles.
The mass and softening lengths of individual gas (dark matter)
particles in the high-resolution region are $5.86\times10^7$
($3.95\times10^8$) ${\rm M}_\odot$ and 2.27 (4.29) kpc, respectively.

The elliptical galaxy found by KG05
\citep[see also][]{kg03b}
has a total virial mass of $2\times10^{13}$~${\rm M_\odot}$.
The galaxy is relatively isolated, with only a few low-mass satellites
remaining at $z=0$. Figure~1 of \citet{kg03b} shows the morphological 
evolution of dark matter in the simulation volume and the evolution 
of the stellar component of the target galaxy. 
The galaxy forms through conventional
hierarchical clustering between redshifts $z$=3 and $z$=1. The morphology
has not changed dramatically since $z$=1. 

KG05 introduced an active galactic nuclei (AGN) heating model
in the simulation, and found that self-regulated activity of
the AGN can suppress significant late-time star formation 
-- characteristics not encountered in
traditional dynamical models of ellipticals. As a result,
their model (model 2 of KG05) succeeds in reproducing both the 
observed X-ray and optical properties of a massive elliptical galaxy.

In the chemodynamical simulation, the simulated star particles
each carry their own age and metallicity ``tag'', which enables us to
generate an optical-to-near infrared spectral energy distribution for the
simulated galaxy, when combined with our population synthesis code 
\citep[which
itself is based upon the population synthesis models of][]{ka97}. 
The population synthesis models are used to construct the photometric 
images used in our analysis.
  
\begin{figure*}
\plotone{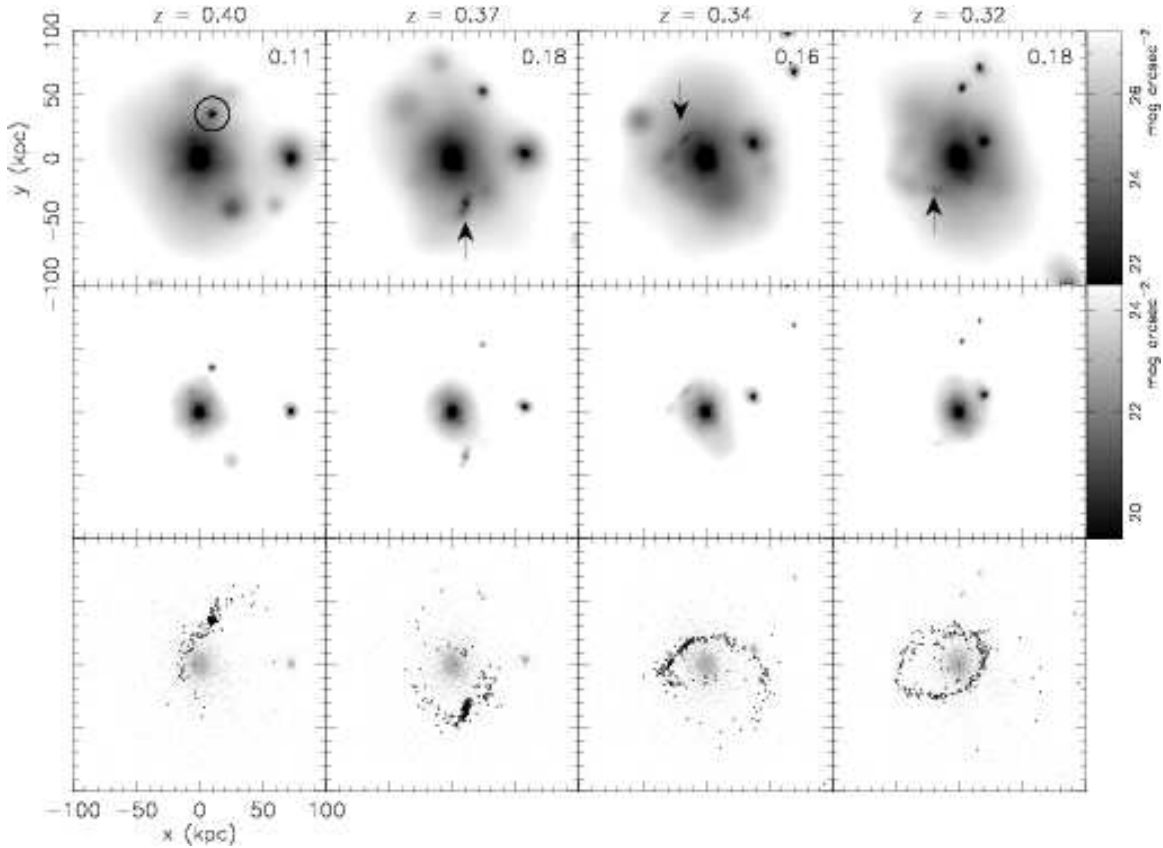}
\caption{
Deep (upper panels) and shallow (middle panels) predicted $R$-band 
images of the target galaxy at the $x-y$ projections
over the redshift range $z=0.4$ to $z=0.32$.
Bottom panels represent the distributions of star particles (gray dots),
the particles within the infalling satellite (see text for details)
are highlighted by black dots.
The infalling satellite galaxy is highlighted by the circle 
in the top-left panel. 
The prominent features are indicated by arrows.
Numbers shown in the top panels are 
the tidal parameter, $t$, (see text) for each image.
\label{fig-z1xy}}
\end{figure*}

\begin{figure*}
\plotone{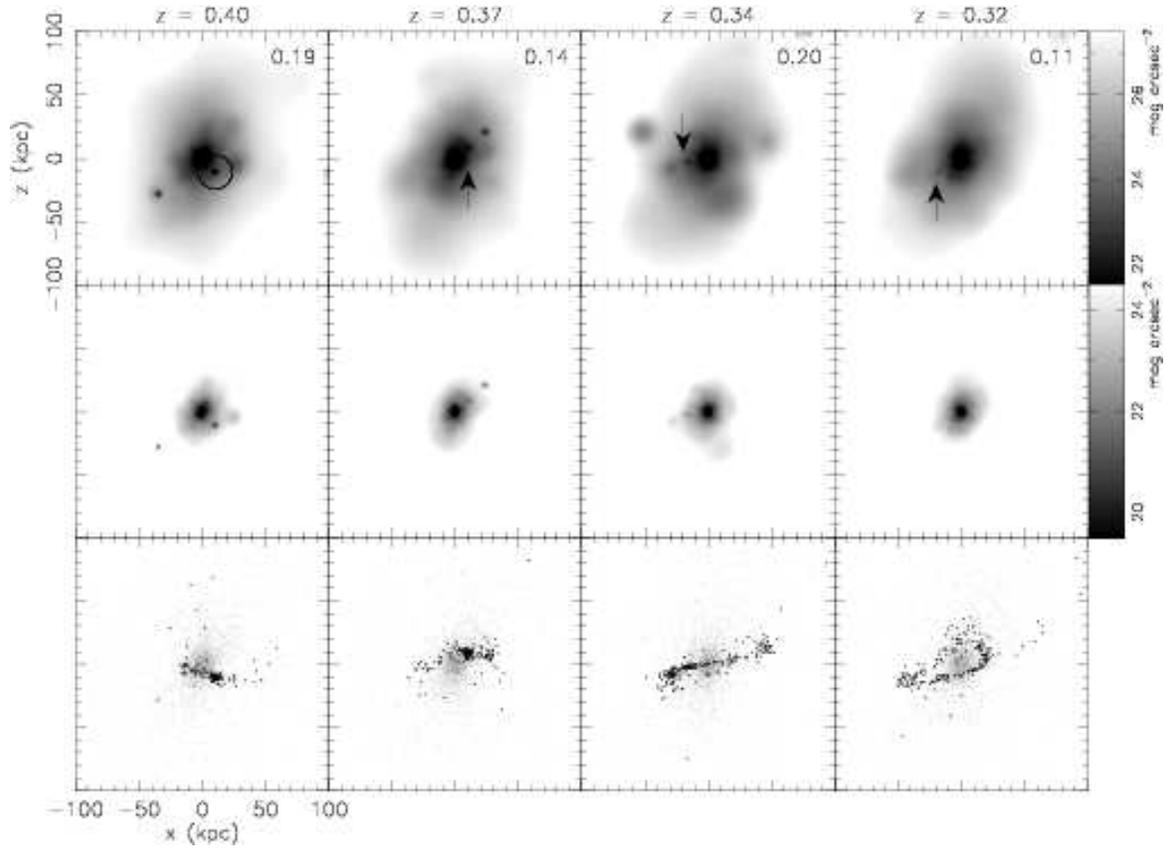}
\caption{Same as Figure \ref{fig-z1xy}, but for the $x-z$ projections.
The high value of the tidal parameter at $z=0.4$ is 
caused by poor model subtraction due to the small separation
between the target galaxy and the infalling satellite.
\label{fig-z1xz}}
\end{figure*}

\begin{figure*}
\plotone{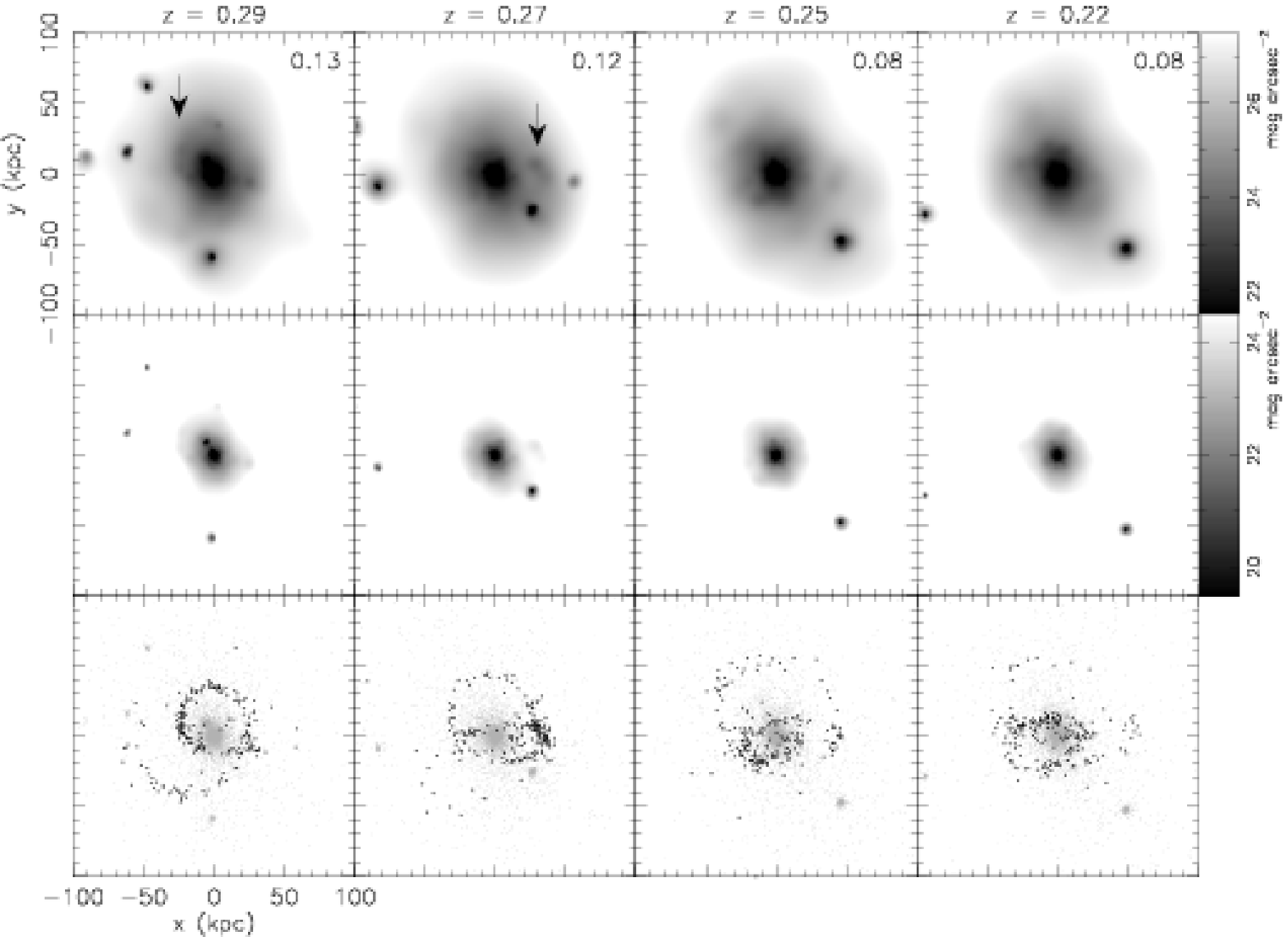}
\caption{Same as Figure \ref{fig-z1xy}, but for redshift range
$z=0.29$ to $z=0.22$.
\label{fig-z2xy}}
\end{figure*}

\begin{figure*}
\plotone{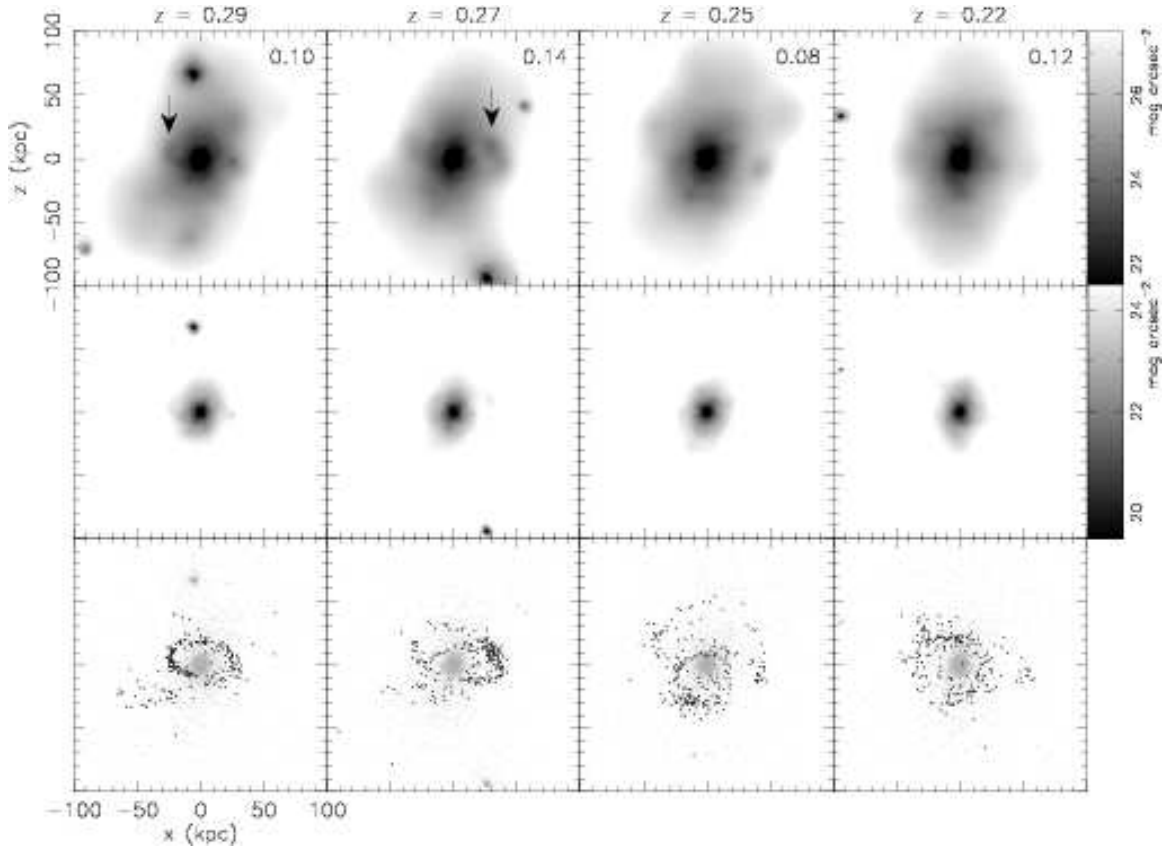}
\caption{Same as Figure \ref{fig-z2xy}, but for the $x-z$ projections.
\label{fig-z2xz}}
\end{figure*}

\begin{figure}
\plotone{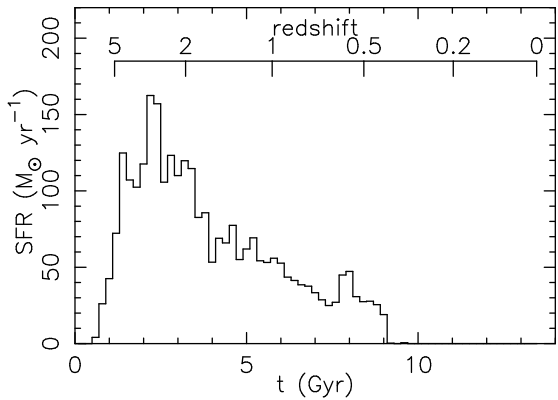}
\caption{
Time-variation of the star formation rate for the target galaxy.
\label{fig-sfr}}
\end{figure}

\begin{figure}
\plotone{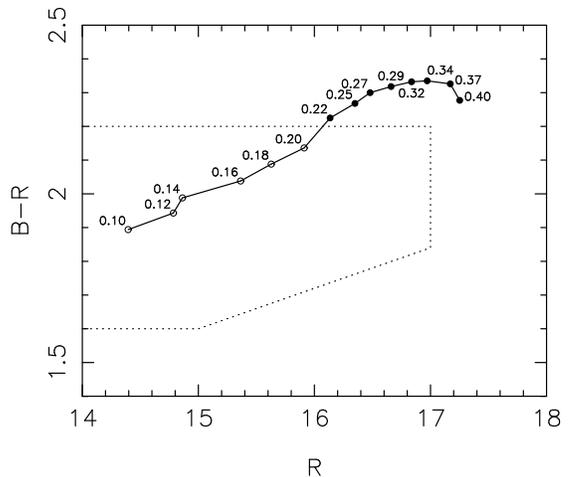}
\caption{
 Time variation of $B-R$ color and the $R$-band magnitude
for the target galaxy. The numbers indicate redshift.
The filled circles are the colors and magnitudes before $z=0.2$,
which we focus on in this paper. The open circles present
the evolution after the stages which are discussed in this paper.
The area enclosed by the dotted lines are the selection criteria
in vD05.
\label{fig-rbrz}}
\end{figure}

\begin{figure*}
 \centering
 \leavevmode
 \includegraphics[width=0.3\linewidth]{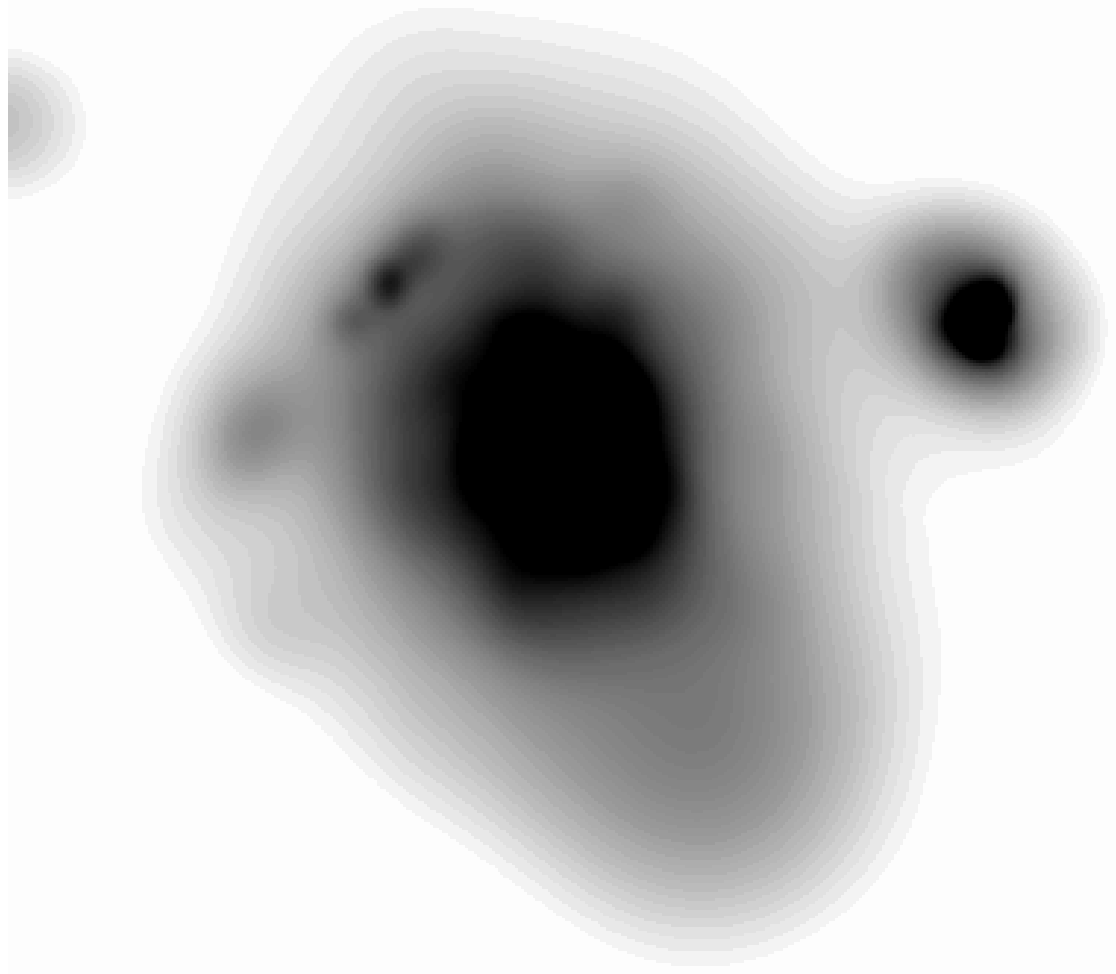}
 \hfil
 \includegraphics[width=0.3\linewidth]{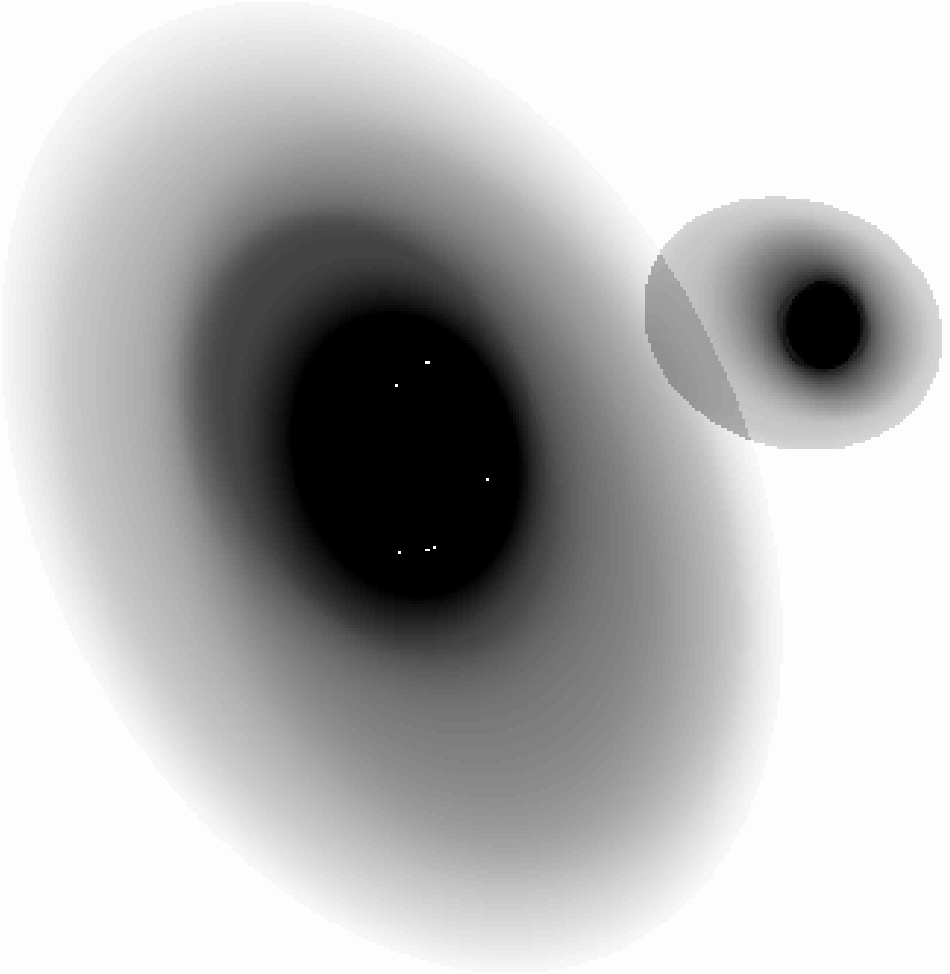}
 \hfil
 \includegraphics[width=0.3\linewidth]{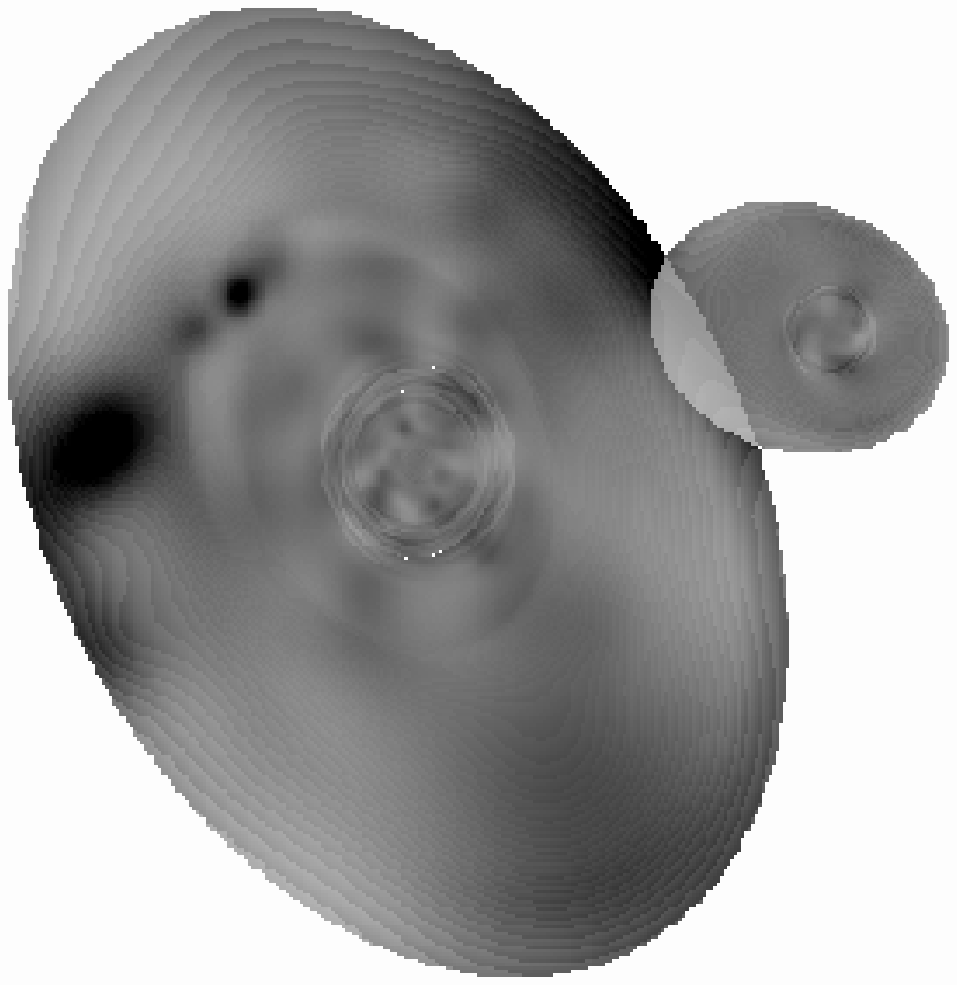}
\caption{Demonstration of the process to derive the value of 
the tidal parameter, $t$ (see text for the details). 
Left panel shows the $R$-band image;
middle panel is the elliptical model; right panel presents
the distortion image.
\label{fig-tproc}}
\end{figure*}

\begin{deluxetable*}{rrrrrrrr}
\tablecolumns{8}
\tablewidth{0pc}
\tablecaption{The luminosity, color, and mass of the target galaxy and the
 infalling satellite at $z=0.4$\tablenotemark{a}.\label{tab-mrm}}
\tablehead{
 \multicolumn{4}{c}{the target galaxy} & \colhead{} &
 \multicolumn{3}{c}{the satellite} \\ \cline{1-4} \cline{6-8} \\
 \colhead{$M_{R}$} & 
 \colhead{$M_R$(age$>$3 Gyr)\tablenotemark{b}}
 & \colhead{$B-R$\tablenotemark{c}} & 
 \colhead{M$_{*}$} & \colhead{} &
 \colhead{$M_{R}$} & \colhead{$B-R$} & \colhead{M$_{*}$} \\
 \colhead{(mag)} & \colhead{(mag)} & \colhead{} & \colhead{($M_{\sun}$)} & 
 \colhead{} & \colhead{(mag)} & \colhead{} & \colhead{($M_{\sun}$)} 
}
\startdata
$-$23.25 & $-$22.50 & 2.53 & $3.1\times10^{11}$ & &
$-$19.17 & 2.39 & $1.0\times10^{10}$ 
\enddata
\tablenotetext{a}{The luminosity is measured within the
 projected diameter of 5 arcsec, i.e. 26.7 kpc at $z=0.4$,
while the mass is measured
 within the three dimentional diameter of 26.7 kpc.}
\tablenotetext{b}{The $R$-band luminosity, when 
 the contribution from the stars younger than 3 Gyr is ignored.}
\tablenotetext{c}{The color, when the contribution from 
 the stars younger than 3 Gyr is ignored.}
\end{deluxetable*}

\section{Dry Tidal Features in the Target Galaxy}
\label{sec-dftg}

Figures \ref{fig-z1xy}-\ref{fig-z2xz} show the morphological
evolution of the target galaxy over the redshift range $z=0.4$
to $z=0.22$ at two different projections: $x-y$ and $x-z$ projection
in our three dimensional simulation coordinate system. 
The top and middle panels in each figure show
the deep ($R<27.0$ mag arcsec$^{-2}$) and
shallow ($R<24.5$ mag arcsec$^{-2}$) images, respectively.
The deep images reveal some small tidal features which are not visible
in the shallow images. The tidal features are similar to 
some of the diffuse tidal tail like features which are observed in
vD05, e.g., 5-994, 9-360, 10-232, 16-650, 16-1302, 17-2819,
21-523, 25-3572 in his sample.
For example, in Figure \ref{fig-z1xy},
the panels at $z=0.37, 0.34$, and 0.32 show faint 
features like tidal tails. In fact, these are
the tidal tails of a small satellite infalling into the target 
galaxy. The infalling satellite galaxy is highlighted by the circle 
in the top-left panel of Figures \ref{fig-z1xy} and \ref{fig-z1xz}.
The bottom panels show the distribution of the
star particles. The grey dots show the whole particles, while
the black dots represent the particles originated from the infalling
satellite. The satellite particles are defined as
the particles whose radius from the center of the satellite
is less than 20 kpc at $z=0.45$,
well before the satellite falls into the target galaxy.
The figures show that the tidal features seen in the deep $R$-band
image correspond to the tidal tails of the infalling satellite.
The tidal tails of the satellite start to be developed 
at $z=0.40$. The satellite is tidally disrupted at $z=0.37$,
and the tidal features disappear around $z=0.25$, which corresponds
to about 1.3 Gyr.
During the period when the tidal tails are visible in the bottom panels, 
the tidal features also appear in the deep image, 
although, not surprisingly, the strength of the features
depends on the projection angle.
vD05 claims that the tidal features appear smooth, showing
little or no evidence for clumps and condensations.
On the other hand, the simulated tidal features suggest the 
appearance of small clump prior to redshift z=0.34, when the
satellite has yet to fully disrupt.  However, one can appreciate 
from Figure 11 of vD05 that some fraction of the observed galaxies 
present  tidal tail-like features accompanied with small clumps, 
for example galaxies 9-360, 10-232, 17-2819, and 21-523. They may 
be an early stage of a minor merger.  After redshift z=0.34, there 
are appear to be no discernible condensations in the tidal tails.

 Another important aspect reported in vD05 is that
both the host galaxy and the tidal features are red, 
and there is no sign of star formation, i.e.,
they are ``dry''. Similarly, in our simulation the produced features
are in fact ``dry tidal features''. 
Figure \ref{fig-sfr} shows the history of the
star formation rate of the system, replotted
from Figure 3 of KG05. 
Star formation stops in the target galaxy
just before $z\sim0.4$, due to the AGN heating (see KG05 for details).
The AGN heating also suppresses star formation during
the infall of the satellite galaxy. 
 
 Table \ref{tab-mrm} displays the luminosity, colors, and 
mass of the target galaxy and the infalling satellite
at $z=0.4$, i.e., before they start merging.
Here, we adopt the diameter of $D=5$ arcsec for a comparison with vD05.
The stellar mass ratio between the satellite galaxy 
and the target galaxy is about 0.03, which indicates that
the infalling satellite is small, i.e., a minor merger.
This indicates that dry tidal features can
result from 
minor mergers.

 Table \ref{tab-mrm} gives two values for the luminosity of 
the target galaxy:
One is the total luminosity, while the second value is the luminosity
when the contribution from stars which are younger than 3 Gyr
is ignored. 
Since star formation stops just before $z=0.4$ in the target galaxy,
the luminosity at $z=0.4$ is affected
by some young stars. As argued in KG05, this late cessation
of star formation in the simulated galaxy is due to
our simple modeling of the AGN heating. In this simulation,
the AGN heating was started at $z=1$, which is likely too late.
If the AGN heating is added at a higher redshift, star formation
is expected to stop earlier.
For real elliptical galaxies, the epoch 
of the termination of star formation is expected to be rather random,
and objects detected as red elliptical galaxies
are likely at a stage well after 
star formation ceases \citep[e.g.,][]{sb98,vdf01}. 
Thus, the luminosity 
when the young stellar population is ignored is more appropriate for objects 
selected to be red and this is the luminosity we consider here.
For such an object, the luminosity difference, $L_1/L_2=10^{(M_{R,1}-M_{R,2})/2.5}$,
between the target galaxy and the satellite galaxy is 0.05,
which is consistent with the lowest luminosity ratios found
in the sample of ongoing mergers in vD05. In fact, those galaxies, 
1-2874 and 5-2345 in vD05, show clear tidal tails, like those seen at $z=0.37$ 
in Figure \ref{fig-z1xy}.
Table \ref{tab-mrm} also shows the colors for both 
the target galaxy and the satellite, which indicates 
that they are both red, and the difference in $B-R$ is 0.14.
This color difference is driven by the difference in metallicity.

 Figure \ref{fig-rbrz} shows the evolution of the luminosity
and colors of the target galaxies. 
The filled circles indicate the colors and magnitudes before $z=0.2$,
when we found the dry tidal features. The open circles represent
the evolution at later stages. 
The area enclosed by the dotted
lines indicates the selection criteria in vD05.
As discussed in vD05, their selection criteria preferably
selects red galaxies at $z<0.2$. 
The epoch we focus on here is a little earlier, and
the target galaxy is redder than the selection criterion adopted by vD05.
However, the epoch of minor mergers is random, 
and such minor mergers are expected to happen often 
in a $\Lambda$CDM cosmology. 
Hence, the results presented here
indicate that if minor mergers happen at $z<0.2$ for elliptical
galaxies, the dry tidal features observed in vD05 can be produced.

 Finally, we quantitatively confirm that the tidal features
shown here are consistent with the features observed in vD05.
To this end, we have analyzed the tidal parameter, $t$, 
which is defined in Section 5.3 in vD05, for 
all the $R$-band images presented in Figures
\ref{fig-z1xy}-\ref{fig-z2xz}.
Here, we briefly explain our procedures to obtain the tidal parameter.
We refer to the simulated galaxy R-band image as ``$G$''. 
First, the galaxy images are fit by an elliptical model using
the {\tt ellipse} task in IRAF. The center position, ellipticity, and position
angle are allowed to vary with radius. In cases where
there are galaxies which overlap with the target galaxy in projection,
the overlapping objects are also fit. The final model image is indicated by ``$M$''.
Finally, a fractional distortion image, ``$F$'', is created by dividing
$G$ by $M$. When the pixel value of $F$ is defined by $F(x,y)$, and the
mean value of the image is $\overline{F}$,
the tidal parameter $t$ which describes the level of distortion
is defined as $t=\overline{F(x,y)-\overline{F}}$. 
Figure \ref{fig-tproc} demonstrates this procedure.

 The values of $t$ are shown in all the top panels in 
Figures \ref{fig-z1xy}-\ref{fig-z2xz}.
We can see that the $R$-band images which have the tidal features
show higher $t$ value ($t>0.1$). 
During the epoch when the tidal features of the satellite galaxy are
visible in the bottom panels ($z>0.25$), $t$ appears to be higher.
As expected, the $t$ values are lower after the tidal features disappear at $z\sim0.25$.
We also find that the value of $t$ depends on projection.
In some cases, the different 
projections give vastly different $t$ values (compare for example
the $x-y$ and $x-z$ panels at $z=0.32$).

The range of $t$ values is consistent with
the weakly (median value of $t$ is 0.13 in vD05) and 
strongly (median $t=0.19$) disturbed galaxies classified in vD05. 
Thus, quantitatively we confirm that the dry tidal features 
seen in our simulated galaxies are consistent with
the dry tidal features seen in vD05.

\section{Discussion and Conclusions}
\label{sec-conc}

 We have demonstrated that a minor merger of a 
satellite galaxy whose stellar mass is $10^{10}$ M$_{\sun}$ with a massive
elliptical galaxies can produce dry tidal features.
In the present simulation, the 
features last about 1.3 Gyr, although the
lifetime of such features
must depend on the orbit of the satellite galaxy.
Traditionally cosmological numerical simulations
of elliptical galaxy formation have suffered from an inability to suppress
late-time star formation in the simulation 
\citep[e.g.,][]{so98,lbk00,ktt03,tbs03,mnse03,kg03b,bms04},
thereby being inconsistent with
the observed red color in bright elliptical galaxies.
However, self-regulated AGN heating, of the sort introduced in 
KG05 aids in the suppression of late-time star formation.
As a result, both the host galaxy and tidal features in our simulation
have the red colors consistent with those observed by vD05.

In a $\Lambda$CDM cosmology, minor mergers as seen in our
simulations are expected to happen much more frequently than 
major mergers. This might explain the high detection
rate of tidal features in vD05.
In addition, minor mergers are consistent with
the red features observed, because mergers with a small galaxy
have a relatively weak effect on the inter-stellar medium (ISM) 
of the host galaxy.
On the other hand, major mergers must have a stronger effect
on the ISM even for elliptical galaxies, where the system is
likely to suffer from radiative cooling. This in turn likely induces star formation,
resulting in blue galaxy colors.
However, some dry tidal features should be produced from major
mergers, as long as these mergers are also "dry".
In fact, ongoing dry major mergers are observed at various
redshifts below z=1 \citep[e.g., vd05;][]{tvf05,bnm06},
and it is known that such major dry mergers can
create tidal features \citep[e.g.,][]{crbps95,bnm06}.
In addition, such major dry mergers do not violate the
scaling relations of elliptical galaxies 
\citep[e.g.,][]{nkb06,bmq05,bmq06}. Our results should not be 
interpreted as degrading the importance of major dry mergers, but 
instead that dry tidal features do not necessarily derive from a 
major dry merger and that a viable alternate pathway for their
appearance may be that of a minor dry merger.  In other words, 
infalling galaxies over a wide range of mass may be responsible 
for producing the observed dry tidal features in elliptical galaxies.
Therefore, considerable caution is needed to use
the dry tidal features for estimates of the major dry merger rate.
Note that vD05 carefully estimates the major dry merger rate,
taking into account only the sample of ongoing dry major mergers.

\citet{jsb01} demonstrate that
if the infalling galaxy was a dwarf spheroidal with the mass of 
$\sim10^8$ M$_{\sun}$, like seen in the Local Group,
the surface brightness of the tidal debris would be 
lower than $\mu_R\sim28$ mag arcsec$^{-2}$ after the dwarf is well-disrupted.
Thus, mergers with the most common low mass galaxies are apparently not sufficient
to produce the features found by vD05. 
Since the infalling galaxy in our simulation has a stellar mass of  
$10^{10}$ M$_{\sun}$, the surface brightness of the resulting tidal
debris is much higher, making the features 
detectable at the depths achieved in vD05. 
Hence, the dry tidal features observed in vD05 are expected to be
end-products of mergers with a system whose stellar mass
is at least $\sim10^{10}$ M$_{\sun}$.
Systematic studies like that of \citet{jsb01}, but with higher mass
satellites, would be useful for determining the mass range
required to reproduce the observed
dry tidal features. 

 The mechanism which creates the dry tidal feature in our simulation
is similar to what is suggested for the formation
of ripples \citep{fs80} and shells \citep{mc80}
observed in nearby elliptical galaxies \citep[e.g.,][]{ss92,cmz01}.
Previous studies 
demonstrate that both a major \citep[e.g.,][]{hs92} and 
minor \citep[e.g.,][]{pq84,dc86,hq88,kn97} merger can create shell
features around giant elliptical galaxies.
vD05 claims that their dry tidal features are different
from the ripples and shells 
seen in nearby elliptical galaxies,
because the dry tidal features look smoother than these other structures. 
However, vD05 also notes that the resolution of his images make it 
difficult to reach strong conclusions about the differences between these
various types of features.
According to the previous systematic studies
of minor mergers \citep[e.g.,][]{hq88}, smoother features are produced
if the infalling galaxy is more of a velocity dispersion supported 
system, i.e., elliptical galaxies. In addition, the tidal features
are sensitive to the infall orbit, and the previous simulations
demonstrate that off-center infall can create large scale 
features around the host galaxy.
The infalling galaxy seen in our simulation has 
an off-center orbit, and thus creates the tidal features
over a large scale, $\sim$50 kpc in radius.

The dry major merger is considered to be an important mechanism
because one major merger is enough to explain the observed 
factor of two increase in stellar mass in bright red galaxies 
since redshift z=1 \citep{bwm04,fww06}.
It is an important question whether or not minor mergers,
such as those described here, can similarly contribute to
the increase in the mass of bright red galaxies.  To check this, 
we compared the stellar mass of the simulated galaxy
within a radius of 100 kpc at z=1 and z=0. As mentioned above,
the simulated galaxy suffers from an excess of ongoing late-time 
star formation (prior to $z\sim0.4$ -- see also Fig. \ref{fig-sfr}).
Hence, we exclude stars whose age at the present epoch is less than 8 Gyrs.
As a result, we find little increase in stellar mass subsequent to
$z=1$ (less than 5\% growth). This is because the simulated galaxy 
does not undergo a major merger subsequent to redshift $z=1.5$, and 
the disrupted satellites are much less massive than the host galaxy. 
(Note that in Table \ref{tab-mrm} we measure the stellar mass within 
a small aperture and therefore underestimate the total stellar mass 
of the host galaxy.)
Therefore, minor mergers such as those described here are unlikely to 
contribute to the observed mass evolution of red galaxies.
This means that major dry mergers must occur for a significant
fraction of the red galaxy population, although we emphasise again
that while potentially the dominant evolutionary pathway, it is not a 
mutually exclusive one. Interestingly, \citet{kda06}
claim that the stellar mass increase is important for the
relatively faint early-type galaxies \citep[see also][]{ar06}.
There may be a mass dependence of the dry major merger rate.

Our simulations of a single elliptical galaxy leaves many 
open questions. For example, how often do minor mergers happen
for bright elliptical galaxies and how do the resulting features depend 
on the mass and morphology of the infalling galaxy. To address these 
questions, simulations of a large sample of ellipticals will be required.

\acknowledgments

DK thanks the financial support of the JSPS, through 
Postdoctoral Fellowship for research abroad. 
The Australian Research Council, through
its Discovery Project scheme, is gratefully acknowledged.
We acknowledge the Yukawa Institute Computer Facility,
the Astronomical Data Analysis Center of the National Astronomical
Observatory, Japan (project ID: wmn14a), the
Institute of Space and Astronautical Science 
of Japan Aerospace Exploration Agency, and
the Australian and Victorian Partnerships for Advanced
Computing, where the numerical computations for this paper were
performed.


\end{document}